\documentstyle[aps,prl,multicol,epsfig]{revtex}
\begin{document}
\draft
\title{
Electron Addition Spectrum in
the Supersymmetric $t$-$J$ Model\\
with Inverse-Square Interaction}
\author{
Mitsuhiro Arikawa,$^1$ Yasuhiro Saiga,$^2$ and
Yoshio
Kuramoto$^3$
}
\address{
$^1$Max-Planck-Institut f\"{u}r Physik komplexer Systeme,N\"{o}thnizer
Str.
38, D-01187 Dresden,
Germany\\
$^2$Institute for Solid State Physics, University of Tokyo, Kashiwa-no-ha
5-1-5, Kashiwa 277-8581, Japan\\
$^3$Department of Physics, Tohoku University, Sendai 980-8578,
Japan
}

\maketitle

\begin{abstract}
The electron addition spectrum $A^+(k,\omega)$ is obtained analytically
for the one-dimensional
(1D) supersymmetric $t$-$J$
model with $1/r^2$ interaction.
The result is obtained first for a small-sized system
and its validity is checked against the numerical calculation.
Then the general expression is found which is valid
for arbitrary size of the system.
The thermodynamic limit of $A^+(k,\omega)$ has a simple analytic form with
contributions from 
one spinon, one holon and one antiholon all of which obey fractional
statistics.
The upper edge of $A^+(k,\omega)$ in the $(k,\omega)$ plane
includes a delta-function peak which reduces to that of the single-electron
band in the low-density limit.

\end{abstract}


%%%%%%%%%
\begin{multicols}{2}

The concept of spinons and holons, both of which obey
the fractional statistics \cite{Haldane}, has turned out to be useful
in approaching to 1D electron systems.
In terms of these quasi-particles one can inquire into
not only the low-energy and low-wavelength limit, but
the global feature of the dynamics.
Hence special interest has been cherished  in the global dynamics from both
theoretical and 
experimental points of view.
For example, angle resolved photoemission \cite{ShenTakahashi} has revealed
some evidence 
of the spin-charge separation by resolution of holon and spinon
contributions.  
On the theoretical side,
numerical studies have been performed for
the 1D {\it t-J} model for
a small number of lattice sites \cite{Tohyama} and some structures have been
ascribed to spinons and
holons.  For deeper understanding of the overall dynamics, demand is
growing for 
analytic theory which can go to the thermodynamic limit.
Partly analytic theory is available for the single-particle spectral
functions 
of the {\it t-J} model in the $J \rightarrow 0$ limit \cite{Penc}.
A notable feature is that a satellite
band is observed whose intensity is comparable to that of the main band.
It is natural to ask how the finite $J$ influences the dynamics.

In the supersymmetric {\it t-J} model with $1/r^2$ interaction
\cite{KY}, spinons and holons appear in the simplest manner.
In fact exact thermodynamics for the model \cite{KK95J} can be
interpreted in terms of free spinons and holons.
Ha and Haldane \cite{HaHaldane} analyzed numerical results for
dynamics in finite-sized systems, and found that only a few number of
elementary excitations
contribute to
spectral functions.
They proposed a momentum-frequency region where
each spectral function takes nonzero values in the thermodynamic limit,
but 
they did not obtain the spectral functions themselves.
Recently, exact results
have been derived 
for a particular component \cite{Kato}, and for a
particular momentum range
of the spectral weight \cite{Ari}.

In this 
paper we report on
the analytical result
of the electron
addition spectrum for the
{\it t-J} model  
at zero temperature.
The electron addition spectral function is relevant to the angle resolved
inverse photoemission
spectroscopy.
Our result constitutes the first analytical knowledge for dynamical
quantities of lattice electrons
with no restriction on the system size, the 
density and the momentum-frequency range.
Although we cannot provide the formal proof for the exactness, the analytic
results
show complete agreement with numerical results for finite systems with
various sizes.
Hence 
our result in
the thermodynamic limit is also expected to be exact.

We consider the supersymmetric $t$-$J$ model given by
\begin{eqnarray}
{\cal H}_{tJ}
&=&
{\cal P}
\sum_{i<j}
\left[-t_{ij}\sum_{\sigma=\uparrow,\downarrow}
\left( c^{\dagger}_{i\sigma}c_{j\sigma} + h.c. \right)
\right.
\nonumber \\
& &
\left.
+J_{ij}
\left(\mbox{\boldmath $S$}_{i}\cdot\mbox{\boldmath $S$}_{j}
-\frac{1}{4} n_{i}n_{j}\right)
\right]
{\cal P},
\label{Tj-Hamiltonian}
\end{eqnarray}
where $c_{i\sigma}$ is the annihilation operator of an electron at
site $i$ with spin
$\sigma$, $n_{i}$ is the number operator and $\mbox{\boldmath
$S$}_{i}$ is the spin
operator.
The projection operator
${\cal P}$
excludes double occupation at each site.
The transfer and exchange energies are given by
$t_{ij}=J_{ij}/2=t D^{-2}_{ij}$
where $D_{ij} = (N/{\pi})\sin \left( \pi\left(i-j\right)/N\right)$
with 
even $N$ 
being the number of lattice 
sites, 
and the lattice constant is
unity.
The electron addition spectral function with the ground state $|0 \rangle$
is defined by
\begin{eqnarray}
A^+ \left(k,\omega \right) &=&\sum_{\nu}
| \langle \nu ; N_{\rm e}+1 | c_{k \sigma}^\dagger | 0 ; N_{\rm e} \rangle
|^2 \nonumber \\
& & \ \times
\delta \left( \omega -E_{\nu}(N_{\rm e}+1)+E_0(N_{\rm e})
+\zeta  \right),
\end{eqnarray}
where $N_{\rm e}$ is the total electron number,
$\zeta$ the chemical potential,
$c_{k \sigma}^\dagger=N^{-1/2}\sum_l c_{l \sigma}^{\dagger} {\boldmath
e}^{{\rm i}k l}$, and
$ | \nu \rangle $ denotes an
eigenstate of the Hamiltonian with energy $E_{\nu}$.

We first give our main result  
and then its derivation.
The addition spectrum includes the dispersion relations $\epsilon_{\rm
s}(q)$ of spinons, 
$\epsilon_{\rm h}(q)$ of holons, and $\epsilon_{\rm a}(q)$ of antiholons.
They are given in units of $t$ by
\cite{KK95J,HaHaldane}
$\epsilon_{\rm s}(q) =  q (v_{\rm s} -q)$,
$\epsilon_{\rm h}(q) = q(v_{\rm c} +q)$
and 
$\epsilon_{\rm a}(q) =  q (2v_{\rm c} -q)/2$,
where $v_{\rm s}=\pi$ and $v_{\rm c}=\pi(1-\bar{n})$
with $\bar{n} = N_{\rm e}/N = 2 k_{\rm F}/\pi$.

Analytical expression of $A^+(k,\omega)$
with $0\le k < 2\pi$
consists of the following components:
\begin{equation}
A^+(k,\omega) =A_{\rm R}(k,\omega)+ A_{\rm L}(k,\omega)+A_{\rm
U}(k,\omega),
\label{AAA}
\end{equation}
where 
\begin{eqnarray}
A_{\rm R}(k,\omega)
&=&
\frac{1}{4\pi}
\int_0^ {k_{\rm F}} d q_{\rm h}
\int_0^{k_{\rm F}-q_{\rm h}}
d q_{\rm s} 
\int_0^{2\pi-4k_{\rm F}} d q_{\rm a}
\nonumber \\
& &
\times
\delta(k-k_{\rm F}-q_{\rm s}-q_{\rm h}-q_{\rm a})
\nonumber \\
& &
\times
\delta(\omega-\epsilon_{\rm s}(q_{\rm s})
-\epsilon_{\rm h}(q_{\rm h})
-\epsilon_{\rm a}(q_{\rm a})
)
\nonumber \\
& &
\times
\frac
{\epsilon_{\rm s}(q_{\rm s})^{g_{\rm s}-1} \epsilon_{\rm h}(q_{\rm
h})^{g_{\rm h}-1}
 \epsilon_{\rm a}(q_{\rm a})^{g_{\rm a}-1}}
 {(q_{\rm h}+q_{\rm a}/2)^2},
\label{kfn}
\end{eqnarray}
and $A_{\rm L}(k,\omega)=A_{\rm R}(2\pi-k, \omega)$.
In 
Eq.(\ref{kfn}) we have $g_{\rm s}=1/2$, $g_{\rm h}=1/2$ and $g_{\rm
a}=2$,
which correspond to statistical parameters of spinons, holons, and
antiholons, respectively \cite
{KK95J,HaHaldane}.
Thus the matrix element in Eq.(\ref{kfn})
can be interpreted in terms of the fractional statistics as in the case of
correlation functions
in related 
continuum systems \cite{KYA}.
The third component is given by
\begin{equation}
A_{\rm U}(k,\omega)  =
\sqrt{
\frac{\epsilon_{\rm a}(k-2k_{\rm F})}{\epsilon_0(k)}
}
\delta(\omega-\omega_{\rm aU}(k)),
\label{kfn2}
\end{equation}
which contributes only in the region $2k_{\rm F} \le k \le 2\pi-2k_{\rm F}$.
Here $\epsilon_0(k) \equiv k (\pi - k/2)$
describes the spectrum of non-interacting electrons,
and
$
\omega_{\rm aU}(k) 
\equiv 
\epsilon_{\rm s}(k_{\rm F}) +\epsilon_{\rm a}(k-2k_{\rm
F}).
$
The $A_{\rm U}(k,\omega)$ can be regarded as contribution from antiholons
with 
fixed-energies of spinons and holons.
In the dilute limit $(k_{\rm F}\rightarrow 0)$,
$\omega_{\rm aU}(k)$ as well as $\epsilon_{\rm a}(k-2k_{\rm F})$
tend to $\epsilon_0(k)$.
Hence the coefficient in Eq.(\ref{kfn2}) becomes unity,
and $A_{\rm U}(k,\omega)$ tends to the spectral intensity of non-interacting
electrons.
The other contributions $A_{\rm R,L}(k,\omega)$
can be neglected in this limit.

Figure 1 shows the spectral edges by solid lines together with the spectral
intensities for a 
finite system 
as explained later.
The threshold behavior in
$A_{\rm R}(k,\omega)$ is derived as follows:
As the frequency approaches an upper edge
given by $\omega_{\rm s}(k)
\equiv 
\epsilon_{\rm s}(k-k_{\rm F})$ with $k_{\rm
F} \le k \le 2k_{\rm F}$,
$A_{\rm R}(k,\omega)$ diverges as
$[\omega_{\rm s}(k)-\omega]^{-1/2}$.
Along the connecting upper edge
for $2k_{\rm F} \le k \le 2\pi -2k_{\rm F}$,
$A_{\rm R}(k,\omega)$ has a stepwise discontinuity
such as $\theta(\omega_{\rm
aU}(k)-\omega)$. 
Here $\theta(x)$ is a
Heaviside step function.
On the other hand, near a lower edge given by
$\omega_{\rm h}(k) 
\equiv 
\epsilon_{\rm h}(k+3k_{\rm F}-2\pi)$ with $2\pi-3k_{\rm
F} \le k \le 2\pi-2k_{\rm
F}$, 
$A_{\rm R}(k,\omega)$ behaves as $[\omega-\omega_{\rm h}(k)]^{3/2}$.
The feature is in contrast to the result of Ref.4
where the spectrum is enhanced near $\omega_{\rm h}(k)$.
Finally near another lower edge given by
$\omega_{\rm aL}(k)
\equiv
\epsilon_{\rm a}(k-k_{\rm F})$
with $k_{\rm F} \le k \le 2\pi-3k_{\rm F}$,
there is a stepwise discontinuity
such as $\theta(\omega-\omega_{\rm aL}(k))$.
We note that the asymptotic behavior of $A^+(k,\omega)$
is fully consistent with the conformal field theory \cite{KY}.

Let us describe derivation of $A^{+}(k,\omega)$.
We represent a state vector $ | \psi \rangle $ in the {\it t-J} model by
%%%%
\begin{equation}
| \psi \rangle = \sum_{   x^{\rm h} ,  x^{\rm s}  }
\psi (  x^{\rm h} ,x^{\rm s}  )
\prod_{i \in  x^{\rm s}  } S_{i}^-
\prod_{j \in  x^{\rm h}  } h_{j}^\dagger | F \rangle.
\end{equation} 
where $ | F \rangle $ is the fully up-polarized state,
$S_{i}^-  = c_{i,\downarrow}^\dagger c_{i,\uparrow}$, and
$h_i^\dagger =  c_{i,\uparrow}$.
The set of coordinates
$ (x^{\rm h},x^{\rm s}) \equiv x$
represents 
$ x  =
(x_0^{\rm h},\cdots,x_{N_{\rm h}}^{\rm h},x_1^{\rm s},\cdots,x_{N_{\rm
s}}^{\rm s})$ 
$= (x_0,x_1, \cdots,x_n)$ with $n=N_{\rm h}+N_{\rm s}$.
Here $x_i^{\rm h}$ is for the $i$-th hole and
$x_j^{\rm s}$ is for the $j$-th down-spin electron.
In order to derive eigenfunctions of the system, it is convenient to use the
Sutherland model 
\cite{Sut} with the
SU(1,1) supersymmetry as an auxiliary \cite{wlc,KK95}.
The merit of using the Sutherland model is that much more is known
about properties of the eigenfunctions than those for the {\it t-J}
model.  
The eigenfunctions with the coupling parameter $\beta=1$ correspond to
%.
such part of the set $\{\psi (  x^{\rm h} ,x^{\rm s}) \}$
that belongs to the
Yangian highest weight states (YHWS) with the SU(2,1) supersymmetry
\cite{Talstra}.
In making the correspondence one restricts the coordinates $x$ to integer
variables. 
It is convenient to introduce the complex coordinate by $z_j = \exp(2 \pi i
x_j /L)$
%.
where $L 
\, (=N)$ is the length of the ring-shaped system.
The YHWS can be expressed by polynomials of $z_j$ where the degree of
each variables lies in the range $[-N/2,N/2]$.
The YHWS form a subset of the Fock space and the other
 states can be generated by successive actions of Yangian generators on the
YHWS. 

%.%%
The ground-state wave function for the {\it t-J} model with $N_{\rm h}+1$
holes
and $N_{\rm s}=N_{\rm e}/2$ down-spin electrons
is given by \cite{KY}
\begin{eqnarray}
\Psi_{\rm GS} 
&=&
\prod_{i \in [0,n] } z_i^{-
\beta 
n/2}
\prod_{i \in I_{\rm s}} z_i^{-(N_{\rm s}-1)/2} \nonumber \\
& & \times
\prod_{i < j \in [0,n]} \left( z_i-z_j \right)
^{\beta}
\prod_{i < j \in I_{\rm s}} \left( z_i-z_j  \right),
\label{gs}
\end{eqnarray}
where we have introduced the interval $ I_{\rm s}=[N_{{\rm h}}+1,N_{{\rm h}}+N_{\rm s}] $
for the suffices $i$ and $j$.
We have assumed that both $N_{\rm s}$ and $N_{\rm h}$ are odd so that the
ground state is non-degenerate.
In the following we always take $\beta=1$.
We fix a hole position at $x^{\rm h}_0=0$ in  $\Psi_{\rm GS}$
which corresponds to adding an up-spin electron at this site.
In terms of 
the ground-state wave function
$\tilde{\Psi}_{\rm GS}$
for the system with $N_{\rm h}$ holes
and $N_{\rm s}$ down-spin electrons,  we
can represent the resultant state by
\begin{eqnarray}
\bigl. \Psi_{\rm GS}
\bigr|_{x_0=0}
& = & 
\sum_{m=0}^n (-1)^m e_m(z)
\tilde{\Psi}_{\rm GS}
,
\label{gs+u}
\end{eqnarray}
where $e_m(z) = \sum_{i_1<i_2<\cdots<i_m \in I} z_{i_1} \cdots z_{i_m}$
is the elementary symmetric function of order $m$.
Note that the interval  $[0,n]$ in Eq.(\ref{gs}) is replaced by 
another interval
$I =[1,n]$ in Eq.(\ref
{gs+u}).

The spectrum of the Sutherland model $\cal H_{\rm Su}$ is conveniently
analyzed with use
of a similarity transformation generated by
${\cal O} =  \prod_{i < j \in I} \left( z_i-z_j\right)^{\beta}
\prod_{i \in I} z_i^{-(\beta n+N_{\rm s}-1)/2}$.
We obtain $\hat{{\cal H}} \equiv {\cal O}^{-1}{\cal H_{\rm Su} O}$ as
follows:
\begin{equation}
\hat{{\cal H}}
 = 
\frac{1}{2}
\left( \frac{2 \pi }{L} \right)^2
\sum_{i=1}^{n}
\left(
\hat{d}_i +  \frac{\beta n}{2}-
\frac{N_{\rm s}-1}{2}
\right)^2, 
\end{equation}
where $\hat{d}_i$ is called the Cherednik-Dunkl operator\cite{Mac}.
It is known that the set $\{ \hat{d}_i\}$ can be diagonalized simultaneously
with
real eigenvalue $\bar{\lambda}_i$ for each $\hat{d}_i$.
The resultant eigenfunctions are polynomials $E_\lambda (z;\beta)$ of
$z_i$ 
and are called the
non-symmetric Jack polynomials\cite{Mac}.
The polynomial $E_\lambda (z;\beta)$ has a
$triangular$ structure with
respect to a certain ordering on the set of $\{ \lambda \}$ \cite{KK95,Mac};
if one expands $E_\lambda (z;\beta)$ in terms of monomials $\Pi_i
z_i^{\nu_i}$,
the expansion coefficient is zero unless  $\lambda \succeq \nu$, where
$\succeq$ 
describes the ordering relation in the set.

Since we are dealing with identical particles,
eigenfunctions
should satisfy the following conditions of the SU(1,1)
supersymmetry:\\
(i) symmetric with respect to exchange of  $z_i^{\rm h}$'s; \\
(ii) anti-symmetric with respect to exchange of  $z_i^{\rm s}$'s.\\
By taking linear combination of $E_\lambda (z;\beta)$, we can
construct another polynomial $K_\lambda (z;\beta)$ with the SU(1,1)
supersymmetry \cite{BF,Dunkl98,YK}.
We specify the set of momenta as
$\left(\lambda^{\rm h},\lambda^{\rm s} \right)
\equiv (\lambda^{\rm h}_1,\cdots, \lambda^{\rm h}_{N_{\rm
h}},\lambda^{\rm s}_1,\cdots,\lambda^{\rm s}_{N_{\rm s}})$
with
$\lambda^{\rm h}_1 \ge \cdots  \ge \lambda^{\rm h}_{N_{{\rm h}}},
\lambda^{\rm s}_1 > \cdots  > \lambda^{\rm s}_{N_{\rm s}}$.
In this way we can parameterize  $K_{\lambda} (z; \beta)$ by using
$\lambda \equiv \left(\lambda^{\rm
h},\lambda^{\rm s} \right)$.
At the ground state with $2N_{\rm s}+1$ electrons
we have $\lambda
=\tilde{\lambda}_{\rm GS} = (\lambda_{\rm GS}^{\rm
h},\lambda_{\rm GS}^{\rm s})$
 with
$
\lambda_{\rm GS}^{\rm h} = ((N_{\rm s}-1)/2, (N_{\rm s}-1)/2, \cdots,
(N_{\rm s}-1)/2)$, and
$\lambda_{\rm GS}^{\rm s} =
(N_{\rm s}-1,N_{\rm s}-2,\cdots,0) $.

%.
We define the inner product $\langle f,g \rangle_0$
for complex functions $f(z)$ and $g(z)$ as
the constant term in the 
Laurent expansion of $f(z)^* g(z)$.
As is clear from the definition of the transformation ${\cal O}$,
$K_\lambda (z;\beta) {\cal O}(z)$ is orthogonal with respect to the above
inner
product.
In order to derive the norm of $K_\lambda (z;\beta)$, we generalize
the procedure taken in Refs.\cite{YK,Takemura,pr}.
The result of lengthy calculation is given by
\begin{equation}
\langle K_{\lambda} {\cal O}, K_{\lambda} {\cal O} \rangle_0
=
N_{{\rm h}} ! N_{\rm s}!
\rho_{\lambda}^{-1}
\langle E_{\lambda} {\cal O}, E_{\lambda} {\cal O} \rangle_0,
\end{equation}
where $\langle E_{\lambda} {\cal O}, E_{\lambda} {\cal O} \rangle_0$ denotes
the norm of 
the
non-symmetric Jack polynomials.
We refer to the literature \cite{Mac} for the explicit form of the
latter 
norm since
it requires many lengthy combinatorial quantities.
The quantity
$\rho_{\lambda} =\rho_{\lambda}^{\rm h}
\rho_{\lambda}^{\rm s}$ is given by
\begin{equation}
\rho_{\lambda}^{\rm h}
=
\prod_{i<j \in I_{\rm h} }
\frac{\bar{\lambda}_i-\bar{\lambda}_j+\beta}
     {\bar{\lambda}_i-\bar{\lambda}_j} ,
\ \
\rho_{\lambda}^{\rm s}
=
\prod_{i<j \in I_{\rm s}}
\frac{\bar{\lambda}_i-\bar{\lambda}_j-\beta}
     {\bar{\lambda}_i-\bar{\lambda}_j},
\end{equation}
where we use the intervals
$I_{\rm s}$ and  $I_{\rm h} =[1,N_{{\rm h}}]$
for the suffices $i$ and $j$.

In order to calculate $A^{+}(k,\omega)$, we need to know
the expansion coefficient $c_{\lambda}$ which appears in
\begin{equation}
e_m (z) \tilde{\Psi}_{\rm GS}
=
\sum_{\lambda}
c_{\lambda} K_{\lambda} (z;\beta)
{\cal O}(z).
\label{K-expansion}
\end{equation}
Using the coefficient $c_{\lambda}$,
$A^{+}(k,\omega)$ can be expressed as
\begin{equation}
A^+(k,\omega) = N_{\rm h}
\sum_\lambda{}^{{\normalsize '}}
\delta(\omega - \Delta E_{\lambda} )
%.
|c_{\lambda}|^2
\frac{ \langle K_{\lambda} {\cal O},  K_{\lambda} {\cal O} \rangle_0}
{ \langle \Psi_{\rm GS}  ,  \Psi_{\rm GS}  \rangle_0 },
\label{formal A+}
\end{equation}
where 
$\Delta E_{\lambda} = E_{\lambda} -E_0 -\zeta $,
the chemical potential $\zeta$ for the finite $N$ is given by
$
\zeta / 
\pi^2 = -(3\bar{n}^2 -6\bar{n} + 4)/12 -(\bar{n}-2)/(2N)+1/(3
N^2),
$ 
 and
the prime means that the summation over $\lambda$ is restricted by the
momentum 
conservation
$ |\lambda|= | \tilde{\lambda}_{\rm GS}| +m$ with
$|\lambda|= \sum_j \lambda_j$ and
$k  = 2 \pi(m + (N_{\rm s}+1)/2)/N $.
The norm ${ \langle \Psi_{\rm GS}  ,  \Psi_{\rm GS}  \rangle_0 }$
can be obtained by a procedure similar to that described above.

From the $triangular$ structure
inherited to $K_\lambda$,
the contributions to the
excited states are limited to
the cases where 
$\lambda^{\rm s}
= \lambda_{\rm GS}^{\rm s} +
(1^{\lambda_{\rm s}},0^{N_{\rm s}-\lambda_{\rm s}})$.
Here $p^\alpha$ means
the 
sequence $p,\cdots,p$
with the number of $p$'s being $\alpha$.
Therefore the spin part can be parameterized
by the single variable $\lambda_{\rm s}$.
We first calculate $K_{\lambda}$ by a brute force
for small systems (up to $N_{\rm h}=5$ and
$N_{\rm s}=15$)
in the small momentum region
$m \le (N_{\rm s}-1)/2$.
In this region, we have $\lambda^{\rm h}
= \lambda_{\rm GS}^{\rm h} + \nu$.
The coefficient $c_{\lambda}$
is zero if the
partition $\nu$ contains
$s=(2,2)$
where $s=(i,j)$ denotes a square in the Young diagram
\cite{Mac}.
We find that the results for $K_{\lambda}$ are expressed
in the following form:
\begin{equation}
c_{\lambda} = 
\prod_{s \in \nu} 
\frac{
-a'(s)+1+\beta' 
(l'(s)-1)}
{a(s) + 1 + \beta' 
l(s)},
\label{c_lambda}
\end{equation}
where 
$\beta' = \beta/(\beta +1) =1/2$,
$a(s) = \nu_i-j$ (arm-length), $a'(s) = j-1$ (arm-colength),
$l(s) = \nu'_j-i$ 
(leg-length) 
with $\nu'_j$ the length of the column,
and $l' (s) = i-1$ (leg-colength).
Thus, we can parameterize $\nu$ as $\nu=(\lambda_{\rm h},1^{\lambda_{\rm
a}-1},0^{N_{\rm h}-\lambda_{\rm a}})$
in 
this small momentum region.
Namely $A^+(k,\omega)$ is
determined by 
the three parameters
$\lambda_{\rm s}$, $\lambda_{\rm h}$ and $\lambda_{\rm a}$, which are
related directly with
momenta of the excitations.
We have checked numerically that
the expression given by Eq.(\ref{c_lambda})  can in fact be extended beyond
the small momentum 
region.
Note that $c_{\lambda}$  is independent of $\lambda_{\rm s}$ which, however,
enters
$A^+(k,\omega)$ through the norm
$\langle K_{\lambda} {\cal O},  K_{\lambda} {\cal O} \rangle_0$.

We thus obtain 
the finite-size version of Eq.(\ref{kfn}):
 \begin{eqnarray}
A_{\rm R} (k,\omega)
&=& 
\sum_\lambda{}^{{\normalsize '}}
I_{\rm R}(\lambda)
\delta(\omega -\Delta E_{\lambda} )
\label{A-finite}
\end{eqnarray}
%.
and similar results with the suffix R replaced by L and U.
In Eq.(\ref{A-finite}) we have
\begin{eqnarray}
I_{\rm R}(\lambda)
&=&
\frac{1}
{ 2\left( \Gamma(1/2) \right)^2 }
\frac{\Gamma(\lambda_{\rm h} -1/2 )\Gamma(\lambda_{\rm h} +N_{\rm h}/2 )}
{ \Gamma( \lambda_{\rm h} )\Gamma( \lambda_{\rm h} + (N_{\rm h}+1)/2)}
 \nonumber \\
 & & \times 
\frac{ \lambda_{\rm a} (N_{\rm h} - \lambda_{\rm a}+1) }
{ ( 2 \lambda_{\rm h} + \lambda_{\rm a} -1 )
 ( 2 \lambda_{\rm h} + \lambda_{\rm a} - 2)} \nonumber \\
 & & \times 
\frac{ \Gamma(\lambda_{\rm s} +1/2 ) }
{ \Gamma( \lambda_{\rm s} +1 ) }
\frac{ \Gamma( N/2 - \lambda_{\rm s}  ) }
{ \Gamma( (N+1)/2 - \lambda_{\rm s}  ) },
\label{finiteversion}
\end{eqnarray}
with 
$Nk/(2\pi)= \lambda_{\rm h} + \lambda_{\rm a} + \lambda_{\rm s} +(N_{\rm
s}-1)/2 $, 
and
\begin{eqnarray}
\lefteqn{\Delta E_{\lambda} =
 \frac{ 2  
 \pi^2 }{ N^2 }
\Bigl[ \lambda_{\rm s} ( N - 1 - 2 \lambda_{\rm s} ) \Bigr. }  \nonumber \\
& & \Bigl. + \lambda_{\rm h} ( 2 \lambda_{\rm h} + N_{\rm h} -2 )
       + (\lambda_{\rm a} -1 ) (N_{\rm h}  -\lambda_{\rm a} ) \Bigr].
       \label{fEn}
\end{eqnarray}
%.
The $triangularity$ leads to
$0 \leq \lambda_{\rm s}+\lambda_{\rm h} \leq (N_{\rm s}-1)/2$.
The parameter $\lambda_{\rm a}$ varies from zero to $N_{\rm h}$.
For the case where $\lambda_{\rm s} > (N_{\rm s}-1)/2$,
we use the reflection symmetry about
$k=\pi$.
%.
We obtain $I_{\rm L}(\lambda) =I_{\rm R}(\bar{\lambda}) $
where the components in
$\bar{\lambda}=(\bar{\lambda}_{\rm s}, \bar{\lambda}_{\rm
h},\bar{\lambda}_{\rm a})$
are given by
$\bar{\lambda}_{\rm s}=N_{\rm s}-\lambda_{\rm s}$, and
by the relation
$
\lambda^{\rm h}
+(0^{N_{\rm h}-\bar{\lambda}_{\rm a}}, 1^{\bar{\lambda}_{\rm
a}-1},\bar{\lambda}_{\rm h})=\lambda_
{\rm GS}^{\rm h}+(1^{N_{\rm h}}).
$
We obtain the allowed range
$0 \leq \bar{\lambda}_{\rm s}+\bar{\lambda}_{\rm h}
\leq (N_{\rm s}+1)/2$.
 
Finally the case where $\lambda_{\rm s}=(N_{\rm s}-1)/2$ should be
considered
separately.
In this case, 
the $triangular$ structure requires that
$c_{\lambda}$ should be unity,
and that 
$\lambda^{\rm h} = \lambda_{\rm GS}^{\rm h}+(1^{\lambda_{\rm a}},0^{N_{\rm
h}-\lambda_{\rm a}})$
with
$1 \leq \lambda_{\rm a} \leq N_{\rm h} $.
We obtain for this special case
\begin{eqnarray}
\lefteqn{I_{\rm U}(\lambda)  =
 \frac{ 
 \Gamma [ (\lambda_{\rm a} +2 )/2  ]
 \Gamma [ (N_{\rm h} - \lambda_{\rm a} +2)/2 ]
 }
 {
 \Gamma [ (\lambda_{\rm a} +1)/2 ]
 \Gamma [ (N_{\rm h} - \lambda_{\rm a} +1)/2 ]
 } }\nonumber \\
 & & \ \times 
 \frac{ 
 \Gamma [ (N_{\rm s} + \lambda_{\rm a} )/2 ]
 \Gamma [ (N_{\rm s} + N_{\rm h} -\lambda_{\rm a}+2)/2 ]
 }
 {
 \Gamma [ (N_{\rm s} + \lambda_{\rm a} +1)/2 ]
 \Gamma [ (N_{\rm s} + N_{\rm h} -\lambda_{\rm a}+2)/2 ]
 }, \label{finiteversion2}
 \end{eqnarray}
 where $N k/(2 \pi)= \lambda_{\rm a} + N_{\rm s} $.
 The excitation energy $\Delta E_{\lambda}$ is given by Eq.(\ref{fEn}) in
the case
 $(\lambda_{\rm s},\lambda_{\rm h},\lambda_{\rm a})=((N_{\rm
s}-1)/2,1,\lambda_{\rm a})$.
 Thus in terms of $I_{\rm R}$, $I_{\rm L}$ and $I_{\rm U}$
we obtain the finite-size version of $A^{+}(k,\omega)$.

\begin{figure}
\epsfig{file=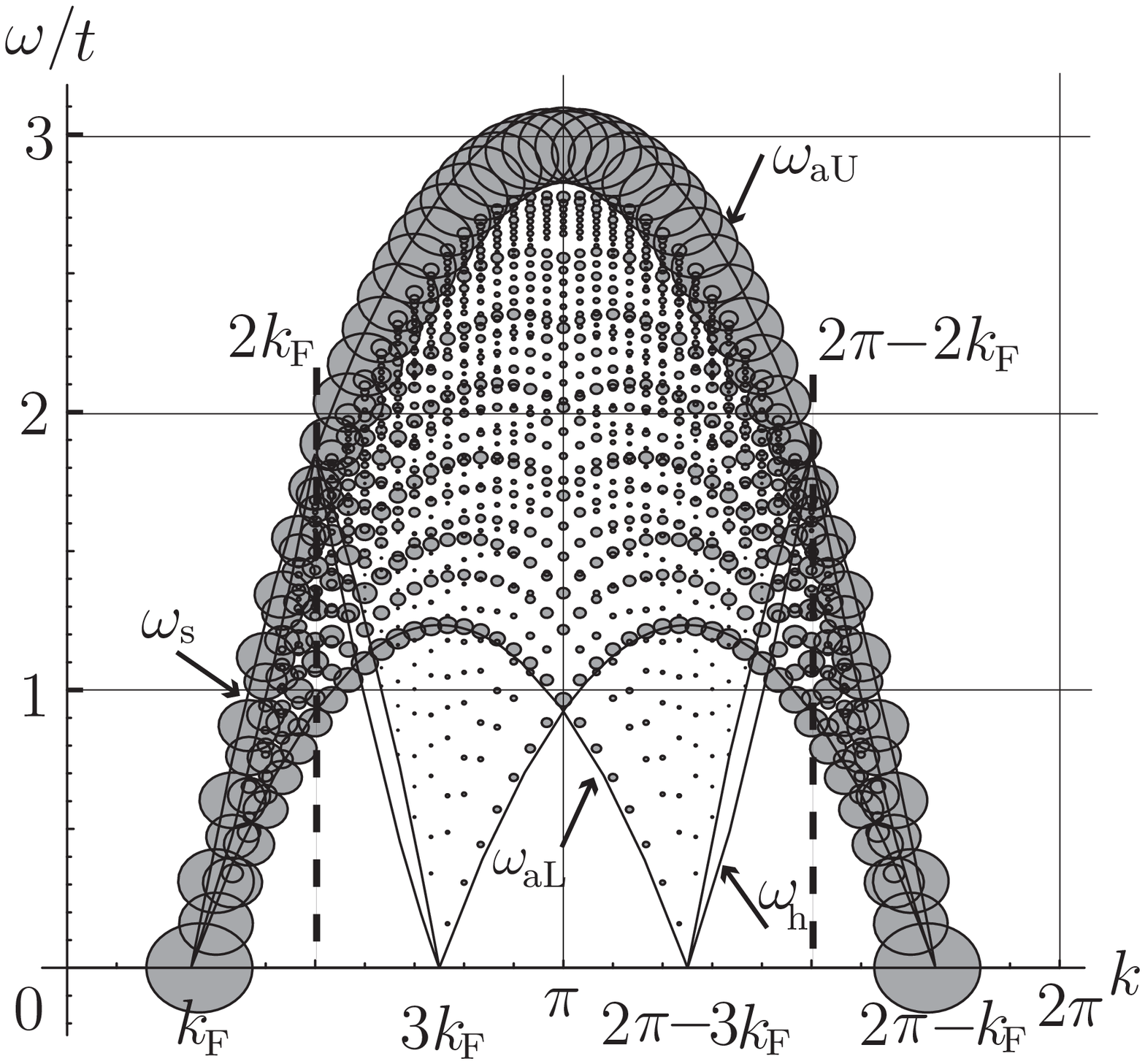, width=8.6cm,height=8cm}
\caption{The electron addition spectrum $A^{+}(k,\omega)$ in the case of
$N=60$, 
$N_{\rm h}=29$ and $N_{\rm s}=15$ with Fermi momentum $k_{\rm F}=\pi/4$.
The intensity is proportional to
the area of 
each oval.  The solid lines are
determined by dispersion relations of the
elementary excitations in the thermodynamic limit.}
\label{maru}
\end{figure}

In Fig. \ref{maru}, we present the result for $N=60$,
$N_{\rm h}=29$ and $N_{\rm s}=15$.
We have checked the validity of
Eqs.(\ref{finiteversion}) and (\ref{finiteversion2})
by comparison with numerical results up to $N=16$ \cite{SK}.
In the special case $\nu = (0^{N_{\rm h}})$,
momenta of the holon and the antiholon are 
both zero, and
we obtain a form different from Eq.(\ref{finiteversion}).
However, this case can be neglected in the thermodynamic limit.
Following the same procedure as that in the spinless Sutherland model
\cite{Ha,Lesage}, we obtain
the expressions (\ref{kfn}) and (\ref{kfn2}) in the thermodynamic limit.

We hope that the quasi-particle structure discussed in this paper is found in future experiment 
of inverse photoemission.

The authors would like to thank Y. Kato and T. Yamamoto for valuable
discussions.
M. A. wishes to thank the support of the Visitor Program of the MPI-PKS.

%
%
%.%%%%%%%%%%%%%
%\begin{center}
%\end{center}
%.%%%%%%%%%%%%%%%%%%%%%%%

\end{multicols}

\end{document}